\documentclass[12pt]{article}
\usepackage[letterpaper,hmargin=1in,vmargin=1.25in]{geometry}
\usepackage{amsmath,amssymb,amsfonts}
\usepackage{algorithm2e}
\newtheorem{theorem}{Theorem}
\newtheorem{example}{Example}
\newcommand{\qed}{\hfill \rule{2mm}{2mm}}

\begin{document}
\title{On Admissible States of Quantum Fourier Transform}
\author{Arpita Maitra and Santanu Sarkar\\
Applied Statistics Unit, Indian Statistical Institute,\\ 
203 B T Road, Kolkata 700108, INDIA\\
Email: arpita76b@rediffmail.com, sarkar.santanu.bir@gmail.com}
\maketitle
\begin{abstract}
We present a general methodology to obtain the basis of qudits which are
admissible to Quantum Fourier Transform (QFT). We first study 
this method for qubits to characterize the ensemble that works for the 
Hadamard transformation (QFT for two dimension). In this regard we identify 
certain incompleteness in the result of Maitra and Parashar (IJQI, 2006). 
Next we characterize the ensemble of qutrits for which QFT is possible. 
Further, some theoretical results related to higher dimensions are also 
discussed.
\end{abstract}
\section{Introduction}
One important quantum gate is the Hadamard gate that has received wide
attention in computer and communication science. There are a number of
seminal papers in quantum computation and information
theory where the Hadamard transform has been used.
The Deutsch-Jozsa algorithm~\cite{qDJ92} to distinguish the constant or 
balanced Boolean functions uses an $n$-dimensional Hadamard gate. Furthermore, 
the Toffoli and Hadamard gates comprise the simplest universal set of quantum
gates~\cite[Chapter 4]{qNC02}. Thus one can easily claim that Hadamard gate 
is one of the most frequently used building blocks in quantum computational 
model. 

An extension of Hadamard transform over higher dimension is the Quantum Fourier
Transform (QFT). QFT has frequent applications in Quantum computation and 
information and one may refer to~\cite[Chapter 5]{qNC02} for detailed 
discussion in this area. The QFT can be seen as linear transformation on 
quantum bits. This is the quantum analogue of the Discrete Fourier 
Transform (DFT). Shor's famous algorithm~\cite{Sh94} for polynomial
time factoring and discrete logarithm are based on Fourier transform which
is a generalization of the Hadamard transform in higher dimension. It has 
applications in other important areas such as quantum phase estimation and 
hidden subgroup problem. It is also important to note that the QFT can be 
performed efficiently on quantum computational framework.

Thus, it is important to identify the quantum states that are admissible
to QFT as those states can be used in a similar manner as the standard
basis and thus states can be used in the same quantum gates that are 
already available. We describe the exact problem in Section~\ref{sprob} little
later. Thus, in this paper we study the universality of QFT.

Pati~\cite{PT02} has proved that one can not design a 
universal Hadamard gate for an arbitrary unknown qubit. This is due to
the simple reason that linearity does not allow
linear superposition of an unknown state $|\psi\rangle$ with its
orthogonal complement $|\psi_{\perp}\rangle$.
Motivated by Pati's work, in~\cite{itx8}, it has been shown how one can
construct a general class of qubit states, for which the
Hadamard gate works as it is. The result of~\cite{itx8} provides certain
ensemble qubit states, for which it is possible to design a universal Hadamard 
gate, are given by $(\alpha + i\beta) |0\rangle + \alpha |1\rangle$. 
In~\cite{itx8}, the orthogonal state of the form 
$b^* |0\rangle - a^* |1\rangle$ has been considered for the state
$a |0\rangle + b |1\rangle$. In fact, all the phase shifts of the state
$b^* |0\rangle - a^* |1\rangle$ are orthogonal to $a |0\rangle + b |1\rangle$.

In this paper, we show that the result of~\cite{itx8} is not a complete
characterization of the qubits such that after application of $U_2$
$|\psi_0\rangle, |\psi_1\rangle$ goes to
$\frac{1}{\sqrt{2}} (|\psi_0\rangle + |\psi_1\rangle)$
and $\frac{1}{\sqrt{2}} (|\psi_0\rangle - |\psi_1\rangle)$ respectively.
We complete the characterization here that is presented in Section~\ref{old2}.

\subsection{Brief Background}
The quantum bits, well known as qubits, can be represented as the 
superposition of $|0\rangle$ and $|1\rangle$ in the form
$|\psi\rangle = \alpha |0\rangle + \beta |1\rangle$,
where $\alpha, \beta$ are complex numbers such that 
$|\alpha|^2 + |\beta|^2 = 1$. The qubits of higher dimensions are called 
qudits. An $n$-dimensional qudit can be represented as
$|\psi_t\rangle = \alpha_{t, 0} |0\rangle + \alpha_{t, 1} |1\rangle + 
\alpha_{t, 2} |2\rangle + \ldots + \alpha_{t, n-1} |n-1\rangle$,
where $\alpha_{t, 0}, \alpha_{t, 1}, \alpha_{t, 2}, \ldots, 
\alpha_{t, n-1}$ are all complex 
numbers and $\sum_{j = 0}^{n-1}|\alpha_{t, j}|^2 = 1$. 
We here index the qudits by $t$ as we will be using more than one qudits at the
same time.

The discrete Fourier transform is usually described as transforming a set 
$x_0, \ldots, x_{n-1}$ of $n$ complex numbers into a set of complex numbers 
$y_0, \ldots, y_{n-1}$ defined by 
$$y_j = U_n(x_j) 
= \frac{1}{\sqrt n} \sum_{k=0}^{n-1} e^{\frac{2\pi i j k}{n}} x_k.$$

The quantum Fourier Transform (QFT) is the counterpart of this transformation
and is defined as follows.
\begin{equation}
\label{qft}
U_n(|j\rangle) = \frac{1}{\sqrt{n}} 
\sum_{k=0}^{n-1} e^{\frac{2\pi i j k}{n}} |k\rangle.
\end{equation}
Quantum Fourier Transform has extremely important role in
Quantum computation as evident from~\cite{qDJ92,Sh94}.
One can write the DFT/QFT matrix $U_n$ as follows for $n$ dimension, when 
$\omega_n = e^{\frac{2\pi i}{n}}$.
$$U_n = \frac{1}{\sqrt{n}}
\left[\begin{array}{crrrrr}
\omega_n^{0\cdot 0} & \omega_n^{0 \cdot 1} & \ldots & \omega_n^{0 \cdot (n-1)}\\
\omega_n^{1\cdot 0} & \omega_n^{1 \cdot 1} & \ldots & \omega_n^{1 \cdot (n-1)}\\
\ldots  & \ldots  & \ldots & \ldots \\
\omega_n^{(n-1)\cdot 0} & \omega_n^{(n-1) \cdot 1} & \ldots & \omega_n^{(n-1) \cdot (n-1)}\\
\end{array}\right].$$ 
Thus DFT/QFT is a unitary transformation given by the unitary matrix $U_n$.
One can view the DFT as a coordinate transformation that specifies the 
components of a vector in a new coordinate system. Thus, given a 
set of qudits $\psi_0, \psi_1, \ldots, \psi_{n-1}$, after application of
QFT, one can get another set of qudits 
$\psi'_0, \psi'_1, \ldots, \psi'_{n-1}$. From the Plancherel 
theorem~\cite{KY68} it is known that the dot product of two vectors is 
preserved under a unitary DFT/QFT transformation. Thus if $\psi_u$ and $\psi_v$
are orthogonal then $\psi'_u$ and $\psi'_v$ will be orthogonal too.

Let us briefly introduce what happens in case of qubits in terms of
Hadamard operations. The Hadamard transform is an example of Fourier transform 
for $n = 2$ and the transformation ($H$ gate) is as follows:
$$U_2 = \frac{1}{\sqrt{2}}
\left[\begin{array}{cr}
1 & 1 \\
1 & -1 \\
\end{array}\right].$$
This takes the orthogonal vectors $|0\rangle$ and $|1\rangle$
to two other orthogonal vectors $\frac{1}{\sqrt{2}} (|0\rangle + |1\rangle)$
and $\frac{1}{\sqrt{2}} (|0\rangle - |1\rangle)$ respectively.
One important question is~\cite{PT02} what is the set of orthogonal vectors
$|\psi_0\rangle, |\psi_1\rangle$ such that after application of
Hadamard gate $H$ one gets two orthogonal vectors
$\frac{1}{\sqrt{2}} (|\psi_0\rangle + |\psi_1\rangle)$
and $\frac{1}{\sqrt{2}} (|\psi_0\rangle - |\psi_1\rangle)$ respectively.
This cannot be true for all the qubits. However,
using linearity, it has been shown~\cite{itx8} that 
$|\psi_0\rangle$ needs to be of the form 
$(\alpha + i\beta) |0\rangle + \alpha |1\rangle$. 

\subsection{The problem}
\label{sprob}
Thus we have the following problem in hand related to QFT.
Consider that an $n$-dimensional qudit can be represented as
$|\psi_t\rangle = \alpha_{t, 0} |0\rangle + \alpha_{t, 1} |1\rangle + 
\alpha_{t, 2} |2\rangle + \ldots + \alpha_{t, n-1} |n-1\rangle$,
where $\alpha_{t, 0}, \alpha_{t, 1}, \alpha_{t, 2}, \ldots, 
\alpha_{t, n-1}$ are all complex 
numbers and $\sum_{j = 0}^{n-1}|\alpha_{t, j}|^2 = 1$. 
We like to characterize the qudits 
$|\psi_0\rangle, |\psi_1\rangle, \ldots, |\psi_{n-1}\rangle$ such that
\begin{equation}
\label{gqft}
U_n(|\psi_j\rangle) = \frac{1}{\sqrt{n}} 
\sum_{k=0}^{n-1} e^{\frac{2\pi i j k}{n}} |\psi_k\rangle.
\end{equation}
It is clear that this is true when 
$|\psi_0\rangle = |0\rangle$, 
$|\psi_1\rangle = |1\rangle$, $\ldots$, 
$|\psi_{n-1}\rangle = |n-1\rangle$. However, it is not true in general
and it is an important theoretical question to characterize such ensembles.

Looking at $U_n$ as a matrix as we have described above, 
$U_n(|\psi_j\rangle)$ can be seen as $U_n \times |\psi_j\rangle$ interpreting 
$|\psi_j\rangle$ as a column vector  
$\left[\begin{array}{c}
\alpha_{j, 0}\\
\alpha_{j, 1}\\  
\ldots\\
\alpha_{j, n-1}
\end{array}\right]$.
Thus, $U_n \times (|\psi_0\rangle, |\psi_1\rangle, \ldots, 
|\psi_{n-1}\rangle)$ can be seen as 
$U_n \times A_n$, where,
$$A_n = \left[\begin{array}{crrr}
\alpha_{0, 0} & \alpha_{1, 0} & \ldots & \alpha_{n-1, 0}\\ 
\alpha_{0, 1} & \alpha_{1, 1} & \ldots & \alpha_{n-1, 1}\\ 
\ldots & \ldots & \ldots & \ldots\\
\alpha_{0, n-1} & \alpha_{1, n-1} & \ldots & \alpha_{n-1, n-1}
\end{array}\right].$$

Now, linearity gives that 
$U(\psi_j) = \alpha_{j, 0} U(|0\rangle) + \alpha_{j, 1} U(|1\rangle) +
\ldots + \alpha_{j, n-1} U(|n-1\rangle)$. From this it is clear to note that
for linearity, we need 
$$U_n A_n = A_n U_n.$$ This provides $n^2$ many constraints
on the elements of the matrix $A_n$ and based on those constraints one can
characterize 
$|\psi_0\rangle, |\psi_1\rangle, \ldots, |\psi_{n-1}\rangle$ that satisfy
Equation~\ref{gqft}.

\subsection{Outline of the paper}
In this paper we point out certain incompleteness in the result of~\cite{itx8} 
and complete the characterization in Section~\ref{old2}.
In Section~\ref{q3}, the characterization related to the qutrits that satisfy 
the QFT are presented. Some brief results related to QFT for quantum states of 
higher dimensions are presented in Section~\ref{sec4} and we explain that 
the nature of the solutions (symmetric or asymmetric) depends on the 
eigenvalues of the QFT matrix. Section~\ref{sec5} concludes the paper.

\section{The case for qubits}
\label{old2}
To study the simplest case, we may revisit the work of~\cite{itx8} in this 
model, which only considers the case $n = 2$. In this case,
\begin{eqnarray}
\label{aa1x}
|\psi_0\rangle & = & \alpha_{0, 0} |0\rangle + \alpha_{0, 1} |1\rangle,\\ 
|\psi_1\rangle & = & \alpha_{1, 0} |0\rangle + \alpha_{1, 1} |1\rangle.
\end{eqnarray}
Thus, we have
$$U_2 = \frac{1}{\sqrt{2}}
\left[\begin{array}{cr}
1 & 1 \\
1 & -1 \\
\end{array}\right], 
\mbox{ and }
A_2 = \left[\begin{array}{cr}
\alpha_{0, 0} & \alpha_{1, 0} \\
\alpha_{0, 1} & \alpha_{1, 1}
\end{array}\right].$$

To elaborate, one needs to satisfy 
\begin{eqnarray}
\label{eqt00}
U_2(|\psi_0\rangle) & = & \frac{1}{\sqrt{2}}
(|\psi_0\rangle + |\psi_1\rangle), \nonumber \\
U_2(|\psi_1\rangle) & = & \frac{1}{\sqrt{2}} (|\psi_0\rangle - |\psi_1\rangle). 
\end{eqnarray}

Now from $U_2A_2 = A_2U_2$, we get $2^2 = 4$ equations and then simple
manipulations provide
$$\alpha_{1, 0} = \alpha_{0, 1} = \frac{\alpha_{0, 0} - \alpha_{1, 1}}{2}.$$
Thus, the general ensemble can be written as 
\begin{eqnarray}
\label{gen2}
|\psi_0\rangle & = & (2\alpha_{0, 1} + \alpha_{1, 1}) |0\rangle + \alpha_{0, 1} |1\rangle, \nonumber\\ 
|\psi_1\rangle & = & \alpha_{0, 1} |0\rangle + \alpha_{1, 1} |1\rangle.
\end{eqnarray}
Hence we get the following result.
\begin{theorem}
Let $|\psi_0\rangle, |\psi_1\rangle$ be the
qubits as described in (\ref{aa1x}). Then they will satisfy (\ref{eqt00})
if and only if they are of the form mentioned in (\ref{gen2}).
\end{theorem}
 
When $|\psi_0\rangle$ and $|\psi_1\rangle$ 
are orthogonal, it has been considered in~\cite{itx8} that 
\begin{eqnarray*}
\alpha_{1, 0} & = & \alpha_{1, 0}^*, \mbox{ i. e., } \alpha_{1, 0} \mbox{ is real},\\ 
\alpha_{1, 1} & = & - (2\alpha_{1, 0} + \alpha_{1, 1})^* = 
- 2\alpha_{1, 0} - \alpha_{1, 1}^* \mbox{ as } \alpha_{1, 0} \mbox{ is real, which gives}\\
\alpha_{1, 1} + \alpha_{1, 1}^* & = & - 2\alpha_{1, 0}, \mbox{ and thus, 
Real}(\alpha_{1, 1}) = - \alpha_{1, 0}.
\end{eqnarray*}
Taking $\alpha_{1, 0} = a$, a real number and 
$\alpha_{1, 1} = a + ib$, where $b$ is real too, one can see that 
$|\psi_0\rangle$ is of the form $(a + ib) |0\rangle + a |1\rangle$ as given
in~\cite{itx8}. If one takes 
\begin{eqnarray*}
\alpha_{1, 0} & = & -\alpha_{1, 0}^*, \mbox{ and}\\
\alpha_{1, 1} & = & (2\alpha_{1, 0} + \alpha_{1, 1})^*,
\end{eqnarray*}
then $\alpha_{1, 0}$ becomes imaginary and imaginary part of $\alpha_{1, 1}$
becomes equal to $-\alpha_{1, 0}$. Thus, 
$|\psi_0\rangle$ is of the form $(a + ib) |0\rangle + ib |1\rangle$.
However, these are not the complete characterization of the ensembles
and thus we refute the following claim of~\cite{itx8}:
``We obtain the most general ensemble of qubits, for which it is possible 
to design a universal Hadamard gate." We present the proper characterization 
in the following analysis.

\subsection{The complete solution for $U_2$ taking
$|\psi_0\rangle, |\psi_1\rangle$ orthogonal}
\label{comp2}
Take $|\psi_0\rangle, |\psi_1\rangle$ as in the form mentioned in (\ref{gen2}).
As they satisfy (\ref{eqt00}), if $|\psi_0\rangle, |\psi_1\rangle$ 
are orthogonal, following Plancherel theorem~\cite{KY68},
$U_2(|\psi_0\rangle), U_2(|\psi_1\rangle)$ will be orthogonal too.

Let us take $\alpha_{1, 1} = \alpha + i \beta$
and $\alpha_{0, 1} = \gamma + i \delta$. Putting these in~(\ref{gen2}),
we have
\begin{eqnarray*}
|\psi_0\rangle & = & \left((\alpha+2\gamma) + i (\beta+2\delta)\right) 
|0\rangle + (\gamma + i \delta) |1\rangle\\
|\psi_1\rangle & = & (\gamma + i\delta) |0\rangle + (\alpha+i\beta) |1\rangle. 
\end{eqnarray*}
Thus, exploiting normality of $|\psi_0\rangle, |\psi_1\rangle$ and the 
orthogonality between $|\psi_0\rangle, |\psi_1\rangle$, 
we get the following kinds of conditions.
\begin{enumerate}
\item {\bf Both $\gamma$ and $\delta$ are zero:} 
$\alpha_{0, 1} = 0$ gives a trivial solution where 
$\gamma=\delta=0$, and $\alpha=\pm \sqrt{1-\beta^2}$. 
That is $|\psi_0\rangle$ has only $|0\rangle$ component and
$|\psi_1\rangle$ has only $|1\rangle$ component.

\item {\bf One of $\gamma$ or $\delta$ is zero:} 

If we take $\gamma = 0$ then from orthogonality, we get $\beta = -\delta$, and
thus $\alpha=\pm \sqrt{1-2\delta^2}$. Hence, $|\psi_0\rangle$ is of the form
$\left(\alpha + i \delta\right) |0\rangle + i \delta |1\rangle$.

Taking $\delta = 0$, from orthogonality, we get $\alpha = -\gamma$, and
thus $\beta=\pm \sqrt{1-2\gamma^2}$. Hence, $|\psi_0\rangle$ is of the form
$\left(\gamma + i \beta\right) |0\rangle + \gamma |1\rangle$. 

This covers the cases considered in~\cite{itx8}.

\item\label{three}{\bf Both $\gamma$ and $\delta$ are non-zero:}

$\alpha = \frac{\delta}{\gamma(\delta^2 + \gamma^2)}
\left(\delta(\delta^2 + \gamma^2)
\pm \gamma \sqrt{\delta^2 + \gamma^2 - 2 (\delta^2 + \gamma^2)^2} \right) - \frac{\delta^2 + \gamma^2}{\gamma}$,

$\beta = - \frac{1}{\delta^2+\gamma^2}\left(\delta(\delta^2 + \gamma^2)
\pm \gamma \sqrt{\delta^2 + \gamma^2 - 2 (\delta^2 + \gamma^2)^2} \right)$.

One may note that $|\alpha_{0, 1}|^2 = \delta^2+\gamma^2$.
\end{enumerate}
Item~\ref{three} is not covered in~\cite{itx8}.
If we put $\delta=\gamma=\frac{1}{2}$ in item~\ref{three}, 
we get $\alpha=\beta=-\frac{1}{2}$. 
Hence we have,
$$|\psi_0\rangle = \frac{1+i}{2} (|0\rangle + |1\rangle), \
|\psi_1\rangle  = \frac{1+i}{2} (|0\rangle - |1\rangle).$$ 
Clearly the $|\psi_0\rangle$ above is not of the form given in~\cite{itx8}.

\section{Characterization for Qutrits}
\label{q3}
We will now consider the case for qutrits. Several quantum systems work
on the qutrits and for example one may refer to~\cite{bb384} for a BB84
like cryptographic protocol. For a qutrit, the QFT can be
seen as follows putting $n = 3$ in~(\ref{qft}).
\begin{equation*}
|j\rangle \rightarrow \frac{1}{\sqrt{3}} 
\sum_{k=0}^{2} e^{\frac{2\pi i j k}{3}} |k\rangle.
\end{equation*}
Denoting the transform as $U_3$, one can write it as:
\begin{eqnarray}
U_3(|0\rangle) & = & \frac{1}{\sqrt{3}} (|0\rangle + |1\rangle + |2\rangle),
\nonumber \\
U_3(|1\rangle) & = & \frac{1}{\sqrt{3}} (|0\rangle + \omega_3 |1\rangle +
\omega_3^2 |2\rangle), \nonumber \\
U_3(|2\rangle) & = & \frac{1}{\sqrt{3}} (|0\rangle + \omega_3^2 |1\rangle +
\omega_3 |2\rangle).
\end{eqnarray}
That is, given $\omega_3 = \sqrt[3]{1}$,
$U_3 = \frac{1}{\sqrt{3}}
\left[\begin{array}{crr}
1 & 1 & 1\\
1 & \omega_3 & \omega_3^2 \\
1 & \omega_3^2 & \omega_3 \\
\end{array}\right]$. 

Now we need
\begin{eqnarray}
\label{eqt1}
U_3(|\psi_0\rangle) & = & \frac{1}{\sqrt{3}}
(|\psi_0\rangle + |\psi_1\rangle + |\psi_2\rangle), \nonumber \\
U_3(|\psi_1\rangle) & = & \frac{1}{\sqrt{3}} (|\psi_0\rangle +
\omega_3 |\psi_1\rangle + \omega_3^2 |\psi_2\rangle), \nonumber \\
U_3(|\psi_2\rangle) & = & \frac{1}{\sqrt{3}} (|\psi_0\rangle +
\omega_3^2 |\psi_1\rangle + \omega_3 |\psi_2\rangle).
\end{eqnarray}

The states $|\psi_0\rangle, |\psi_1\rangle, |\psi_2\rangle$ are as follows.
\begin{eqnarray}
\label{aaeq}
|\psi_0\rangle & = & \alpha_{0, 0} |0\rangle + \alpha_{0, 1} |1\rangle + \alpha_{0, 2} |2\rangle \nonumber \\
|\psi_1\rangle & = & \alpha_{1, 0} |0\rangle + \alpha_{1, 1} |1\rangle 
+ \alpha_{1, 2} |2\rangle \nonumber \\
|\psi_2\rangle & = & \alpha_{2, 0} |0\rangle + \alpha_{2, 1} |1\rangle + \alpha_{2, 2} |2\rangle.
\end{eqnarray}

From $U_3 A_3 = A_3 U_3$, we get the following equations.
\begin{eqnarray}
\label{t1} \alpha_{0, 0} + \alpha_{0, 1} + \alpha_{0, 2} & = &  \alpha_{0, 0} + \alpha_{1, 0} + \alpha_{2, 0}\\
\label{t2} \alpha_{0, 0} + \omega_3 \alpha_{0, 1} + \omega_3^2 \alpha_{0, 2} & = &  \alpha_{0, 1} + \alpha_{1, 1} + \alpha_{2, 1}\\
\label{t3} \alpha_{0, 0} + \omega_3^2 \alpha_{0, 1} + \omega_3 \alpha_{0, 2} & = &  \alpha_{0, 2} + \alpha_{1, 2} + \alpha_{2, 2}\\
\label{t4} \alpha_{1, 0} + \alpha_{1, 1} + \alpha_{1, 2} & = &  \alpha_{0, 0} + \omega_3 \alpha_{1, 0} + \omega_3^2 \alpha_{2, 0}\\
\label{t5} \alpha_{1, 0} + \omega_3 \alpha_{1, 1} + \omega_3^2 \alpha_{1, 2} & = &  \alpha_{0, 1} + \omega_3 \alpha_{1, 1} + \omega_3^2 \alpha_{2, 1}\\
\label{t6} \alpha_{1, 0} + \omega_3^2 \alpha_{1, 1} + \omega_3 \alpha_{1, 2} & = &  \alpha_{0, 2} + \omega_3 \alpha_{1, 2} + \omega_3^2 \alpha_{2, 2}\\
\label{t7} \alpha_{2, 0} + \alpha_{2, 1} + \alpha_{2, 2} & = &  \alpha_{0, 0} + \omega_3^2 \alpha_{1, 0} + \omega_3 \alpha_{2, 0}\\
\label{t8} \alpha_{2, 0} + \omega_3 \alpha_{2, 1} + \omega_3^2 \alpha_{2, 2} & = &  \alpha_{0, 1} + \omega_3^2 \alpha_{1, 1} + \omega_3 \alpha_{2, 1}\\
\label{t9} \alpha_{2, 0} + \omega_3^2 \alpha_{2, 1} + \omega_3 \alpha_{2, 2} & = &  \alpha_{0, 2} + \omega_3^2 \alpha_{1, 2} + \omega_3 \alpha_{2, 2}
\end{eqnarray}
From~(\ref{t1}), 
$$\alpha_{0, 1} = \alpha_{1, 0} + \alpha_{2, 0} - \alpha_{0, 2}.$$
Adding~(\ref{t2}) and~(\ref{t3}) and putting the value of $\alpha_{0, 1}$, 
we get $$\alpha_{0, 0} = \alpha_{1, 0} + \alpha_{2, 0} + 
\frac{1}{2} (\alpha_{1, 1} + \alpha_{2, 1} + \alpha_{1, 2} + \alpha_{2, 2}).$$
Then putting both the values of $\alpha_{0, 0}, \alpha_{0, 1}$ in~(\ref{t3}),
one can get 
$$\alpha_{0, 2} = 
\frac{1}{2}(\alpha_{1, 0} + \alpha_{2, 0}) +
\frac{1}{4}\omega_3^2(\alpha_{1, 2} + \alpha_{2, 2}) -
\frac{1}{4}\omega_3^2(\alpha_{1, 1} + \alpha_{2, 1}).$$
Further, replacing $\alpha_{0, 0}$ in~(\ref{t4}), it can be seen that 
$$\alpha_{1, 0} = 
\frac{1}{2}\omega_3^2(\alpha_{1, 1} + \alpha_{1, 2}) -
\frac{1}{2}\omega_3^2(\alpha_{2, 1} + \alpha_{2, 2}) + \alpha_{2, 0}.$$
Now~(\ref{t5}) gives, 
$$\alpha_{1, 0} + \omega_3^2 \alpha_{1, 2} = \alpha_{0, 1} 
+ \omega_3^2 \alpha_{2, 1}$$ and from this it follows that 
$$\alpha_{1, 2} = \alpha_{2, 1}.$$

Similarly from~(\ref{t6}), we get 
$$\alpha_{1, 0} + \omega_3^2 \alpha_{1, 1} = \alpha_{0, 2} +
\omega_3^2 \alpha_{2, 2}.$$ Replacing the values of 
$\alpha_{0, 2}, \alpha_{1, 0}$, we get $$\alpha_{1, 1} = \alpha_{2, 2}.$$
Next from~(\ref{t7}), one can get 
$$\alpha_{2, 0} = \alpha_{1, 0}.$$
Thus, we finally get 
$$\alpha_{0, 1} = \alpha_{1, 0} = \alpha_{2, 0} = \alpha_{0, 2}.$$
Manipulating~(\ref{t8}),~(\ref{t9}), one can check that 
$$\alpha_{2, 1} = \alpha_{1, 2} \mbox{ and } \alpha_{2, 2} = \alpha_{1, 1}.$$

So $|\psi_0\rangle, |\psi_1\rangle, |\psi_2\rangle$ are of the following form.
\begin{eqnarray}
\label{eqt1a}
|\psi_0\rangle & = & (\alpha_{1, 1} + \alpha_{1, 2} + 2\alpha_{0, 1}) |0\rangle + \alpha_{0, 1} |1\rangle + \alpha_{0, 1} |2\rangle), \nonumber \\
|\psi_1\rangle & = & \alpha_{0, 1} |0\rangle + \alpha_{1, 1} |1\rangle + \alpha_{1, 2} |2\rangle), \nonumber \\
|\psi_2\rangle & = & \alpha_{0, 1} |0\rangle + \alpha_{1, 2} |1\rangle + \alpha_{1, 1} |2\rangle).
\end{eqnarray}

Thus we get the following important result.
\begin{theorem}
\label{th2}
Let $|\psi_0\rangle, |\psi_1\rangle, |\psi_2\rangle$ be the
qutrits as described in (\ref{aaeq}). Then they will satisfy (\ref{eqt1})
if and only if they are of the form mentioned in (\ref{eqt1a}).
\end{theorem}

\subsection{Solutions when $|\psi_0\rangle, |\psi_1\rangle, |\psi_2\rangle$ 
are orthogonal}
Let $\alpha_{0, 1} = x_0+iy_0, \alpha_{1, 1} = x_1+iy_1, 
\alpha_{1, 2} = x_2+iy_2$.
Now we will try to obtain relations following~(\ref{eqt1a}). From orthogonality
and normality, we get the following conditions:
\begin{eqnarray}
\label{eq44}
f_1 & = & x_0^2+y_0^2+x_1^2+y_1^2+x_2^2+y_2^2-1 = 0, \nonumber \\
f_2 & = & (x_1+x_2+2x_0)^2+(y_1+y_2+2y_0)^2+2(x_0^2+y_0^2)-1 = 0, \nonumber \\
f_3 & = & (x_1+x_2+2x_0)x_0+(y_1+y_2+2y_0)y_0+x_0x_1+y_0y_1+x_0x_2+y_0y_2 = 0, \nonumber \\
f_4 & = & x_0^2+y_0^2+2x_1x_2+2y_1y_2 = 0,
\end{eqnarray}
over the variables $x_0, y_0, x_1, y_1, x_2, y_2$.
That is we need to find common roots of $f_1,f_2, f_3, f_4$.
Let, $I$ be the ideal generated by $f_1, f_2, f_3, f_4$ over the polynomial 
ring $\mathbb{R}[x_0, x_1, x_2, y_0, y_1, y_2]$. As dimension
of $I$ is 3 (checked using SAGE~\cite{sage}), one can choose any three 
of $x_0, y_0, x_1, y_1$, $x_2, y_2$ and then obtain the values of other three
by putting the chosen in the equations above. 

As an example, one can choose the values of $y_0, y_1, y_2$ and then
try to find $x_0, x_1, x_2$ in terms of $y_0, y_1, y_2$. However, it is 
extremely tedious to write the complete expression. 
Let us first provide an example with some numerical value.
\begin{example}
\label{ex1}
Take $y_0 = y_1 = y_2 = \frac{1}{10}$. Thus we get the following that provide
solutions to~(\ref{eq44}):\\
$x_0 = \sqrt{\frac{11}{75} + \frac{\sqrt{1909}}{300}}$,\\
$x_1 = \frac{1}{180}(90 + 2 \sqrt{3} (44 + \sqrt{1909})^{\frac{3}{2}} -
  135 \sqrt{3 (44 + \sqrt{1909})} - \sqrt{5727 (44 + \sqrt{1909})})$,\\
$x_2 = \frac{1}{180}(-47\sqrt{3 (44 + \sqrt{1909})} + \sqrt{5727(44 +
\sqrt{1909})}$\\
$- \sqrt{2862 + 54 \sqrt{1909} + 12354 (44 + \sqrt{1909}) - 
282 \sqrt{1909} (44 + \sqrt{1909})})$.
\end{example}
Next we carefully study several interesting situations that provide compact 
expressions.

\subsubsection{The solutions when $x_0 = y_0 = 0$}
Given $x_0 = y_0 = 0$, we have the following solutions.
\begin{enumerate}
\item 
$x_1=\pm \frac{\left(y_1^2 +y_2^2 - 1
-\sqrt{(y_1^2+y_2^2-1)^2-4y_1^2y_2^2}\right)
\left(\sqrt{1-y_1^2-y_2^2-\sqrt{(y_1^2+y_2^2-1)^2-4y_1^2y_2^2}}\right)}{2\sqrt{2}y_1y_2}$\\
$x_2=\pm\sqrt{\frac{1-y_1^2-y_2^2-\sqrt{(y_1^2+y_2^2-1)^2-4y_1^2y_2^2}}{2}}$

\item 
$x_1=\pm\frac{\left({y_1^2 + y_2^2 - 1 + \sqrt{(y_1^2+y_2^2-1)^2
-4y_1^2y_2^2}}\right)\left(\sqrt{{1-y_1^2
-y_2^2+\sqrt{(y_1^2+y_2^2-1)^2-4y_1^2y_2^2}}}\right)}{2\sqrt{2}y_1y_2}$\\
$x_2=\pm\sqrt{\frac{1-y_1^2-y_2^2+\sqrt{(y_1^2+y_2^2-1)^2-4y_1^2y_2^2}}{2}}$
\end{enumerate}

In this case, $\alpha_{0, 1} = 0$ and putting that in~(\ref{eqt1a}),
$|\psi_0\rangle, |\psi_1\rangle, |\psi_2\rangle$ are of the following form.
\begin{eqnarray}
\label{two3}
|\psi_0\rangle & = & (\alpha_{1, 1} + \alpha_{1, 2}) |0\rangle, \nonumber \\
|\psi_1\rangle & = & \alpha_{1, 1} |1\rangle + \alpha_{1, 2} |2\rangle), \nonumber \\
|\psi_2\rangle & = & \alpha_{1, 2} |1\rangle + \alpha_{1, 1} |2\rangle).
\end{eqnarray}

Two examples of such states are 
$$|\psi_0\rangle = |0\rangle, \
|\psi_1\rangle = \frac{1}{2}(1-i) |1\rangle + \frac{1}{2} (1+i) |2\rangle, \
|\psi_2\rangle = \frac{1}{2}(1+i) |1\rangle + \frac{1}{2} (1-i) |2\rangle;$$
and
$$|\psi_0\rangle = \frac{i-1}{\sqrt{2}} |0\rangle, \
|\psi_1\rangle = \frac{1}{\sqrt{2}}(-|1\rangle +i |2\rangle), \
|\psi_2\rangle = \frac{1}{\sqrt{2}}(i|1\rangle - |2\rangle).$$

\subsubsection{The solution when $y_0 = 0$, but $x_0 \neq 0$}
\label{nxzy}
When $y_0 =0$ we have following solutions.
\begin{enumerate}
\item \begin{itemize}
$x_0 = \pm\sqrt{\frac{1-(y_1+y_2)^2}{3}}$\\
$x_1 = \mp\frac{\sqrt{1-(y_1+y_2)^2}}{2\sqrt{3}}-\frac{\sqrt{1-(y_1-y_2)^2}}{2}$\\
$x_2 = \mp\frac{\sqrt{1-(y_1+y_2)^2}}{2\sqrt{3}} +\frac{\sqrt{1-(y_1-y_2)^2}}{2}$
\end{itemize}

\item \begin{itemize}
$x_0 = \pm\sqrt{\frac{1-(y_1+y_2)^2}{3}}$\\
$x_1 = \mp\frac{\sqrt{1-(y_1+y_2)^2}}{2\sqrt{3}}+\frac{\sqrt{1-(y_1-y_2)^2}}{2}$\\
$x_2 = \mp\frac{\sqrt{1-(y_1+y_2)^2}}{2\sqrt{3}} -\frac{\sqrt{1-(y_1-y_2)^2}}{2}$
\end{itemize}
\end{enumerate}

\subsubsection{The real solutions}
One interesting situation is when 
$\alpha_{0, 1}, \alpha_{1, 1}, \alpha_{1, 2}$ are all real,
i. e., $y_0 = y_1 = y_2 = 0$. This follows putting $y_1 = y_2 = 0$
in the results of previous section (Section~\ref{nxzy}). One may note that
there are exactly four solutions for this. 
\begin{enumerate}
\item 
$x_0 = \pm \frac{1}{\sqrt{3}}$,
$x_1 = \mp \frac{1}{2\sqrt{3}} -\frac{1}{2}$,
$x_2 = \mp \frac{1}{2\sqrt{3}} +\frac{1}{2}$,

\item 
$x_0 = \pm \frac{1}{\sqrt{3}}$,
$x_1 = \mp \frac{1}{2\sqrt{3}} + \frac{1}{2}$,
$x_2 = \mp \frac{1}{2\sqrt{3}} -\frac{1}{2}$,
\end{enumerate}

Following~(\ref{eq44}) and taking $y_0 = y_1 = y_2 = 0$, one can consider this
as obtaining points of intersection of the following four planes in three 
dimension.
\begin{eqnarray*}
\label{eq441}
x_0^2 + x_1^2+ x_2^2 -1 & = & 0, \nonumber \\
(x_1+x_2+2x_0)^2 + 2x_0^2 - 1 & = & 0, \nonumber \\
(x_1+x_2+2x_0)x_0 + x_0x_1 + x_0x_2 & = & 0, \nonumber \\
x_0^2 + 2x_1x_2 & = & 0.
\end{eqnarray*}

\section{Brief study of the general case}
\label{sec4}
From the previous two sections, it is clear that $A_2, A_3$ are symmetric.
Thus it requires an understanding what happens for the general case.
In this direction, let us first present the following result.
\begin{theorem}
If the Eigen values of $U_{n}$ are distinct, then $A_n$ is symmetric. 
\end{theorem}
{\bf Proof: }
Since the Eigen values of $U_n$ are distinct, $U_n$ is 
diagnosable. Let $T_n$ be a matrix so that $T_n U_n T_n^{-1}$ is diagonal, 
with distinct diagonal entries $u_1, u_2,\ldots,u_n$. Consider a matrix $B_n$
such that $B_nU_n=U_nB_n$.

Now,
$(T_n B_n T_n^{-1})(T_n U_n T_n^{-1}) = (T_n U_n T_n^{-1})(T_n B_n T_n^{-1})$. 
Let, $M_n = T_nB_nT_n^{-1}$. Now $(i, j)$-th entry of 
$M_n (T_nU_nT_n^{-1})$ is $m_{ij} u_j$. 
Also $(i, j)$-th entry of the matrix $(T_nU_nT_n^{-1}) M_n$ is
$u_i m_{ij}$. If $i \neq j$, given $u_i, u_j$ are distinct,
$m_{ij}u_j = u_im_{ij}$ holds iff $m_{ij} = 0$.
Thus, a matrix commuting with $T_n U_n T_n^{-1}$ is diagonal.
Using  interpolation one can find a  polynomial 
$P$ so that $P(T_nU_nT_n^{-1}) = T_n B_n T_n^{-1}$ is that other polynomial. 
Since $P(T_n U_n T_n^{-1}) = T_n \cdot P(U_n)\cdot T_n^{-1}$, 
so $B_n = P(U_n)$. 

Since $U_n$ is symmetric and $B_n$ is a polynomial in $U_n$, $B_n$ will be 
symmetric too. From definition, we have $A_nU_n = U_nA_n$. Thus $A_n$
is symmetric when the Eigen values of $U_n$ are distinct. \qed

The Eigen values of $U_2$ are $\pm 1$ and the Eigen values of $U_3$ are 
$\pm\sqrt{3}, -1+2\omega_3$. Thus, the Eigen values of $U_2$ and $U_3$ are 
distinct and thus $A_2, A_3$ are symmetric as we
have already observed in the previous sections. Now let us look at $U_4$
and $A_4$ which are of the following form:

$$U_4 = \frac{1}{2}
\left[\begin{array}{rrrr}
1 & 1 & 1 &1\\
1 & \omega_4 & \omega_4^2 &\omega_4^3 \\
1 & \omega_4^2 & 1& \omega_4^2 \\
1 & \omega_4^3 & \omega_4^2& \omega_4 \\
\end{array}\right], \
A_4 = \left[\begin{array}{rrrr}
\alpha_{0,0} & \alpha_{1,0} & \alpha_{2,0} &\alpha_{3,0}\\
\alpha_{0,1} & \alpha_{1,1} & \alpha_{2,1} &\alpha_{3,1} \\
\alpha_{0,2} & \alpha_{1,2} &  \alpha_{2,2}& \alpha_{3,2} \\
\alpha_{0,3} & \alpha_{1,3} & \alpha_{2,3}& \alpha_{3,3} \\
\end{array}\right].$$
One may note that $\omega_4 = i$. The Eigen values of $U_4$ are $1, 1,-1, i$, 
which are not distinct and we find that the $A_4$ is indeed not symmetric.

As before, We consider the qudits of the following form.
\begin{eqnarray}
\label{aaeq44}
|\psi_0\rangle & = & \alpha_{0,0} |0\rangle + \alpha_{0,1} |1\rangle + \alpha_{0,2} |2\rangle + \alpha_{0,3} |3\rangle, \nonumber \\
|\psi_1\rangle & = & \alpha_{1,0} |0\rangle + \alpha_{1,1} |1\rangle + \alpha_{1,2} |2\rangle + \alpha_{1,3} |3\rangle, \nonumber \\
|\psi_2\rangle & = & \alpha_{2,0} |0\rangle + \alpha_{2,1} |1\rangle + \alpha_{2,2} |2\rangle + \alpha_{2,3} |3\rangle, \nonumber \\
|\psi_3\rangle & = & \alpha_{3,0} |0\rangle + \alpha_{3,1} |1\rangle + \alpha_{3,2} |2\rangle + \alpha_{3,3} |3\rangle.
\end{eqnarray}

To satisfy QFT, we need
\begin{eqnarray}
\label{eqt144}
U(|\psi_0\rangle) & = & \frac{1}{2}
\left(|\psi_0\rangle + |\psi_1\rangle + |\psi_2\rangle + |\psi_3\rangle\right), \nonumber \\
U(|\psi_1\rangle) & = & \frac{1}{2} \left(|\psi_0\rangle +
\omega_4 |\psi_1\rangle + \omega_4^2 |\psi_2\rangle + \omega_4^3 |\psi_3\rangle\right), \nonumber \\
U(|\psi_2\rangle) & = & \frac{1}{{2}} \left(|\psi_0\rangle +
\omega_4^2 |\psi_1\rangle + |\psi_2\rangle +\omega_4^2 |\psi_3\rangle\right),\nonumber\\
U(|\psi_3\rangle) & = & \frac{1}{{2}} \left(|\psi_0\rangle +
\omega_4^3 |\psi_1\rangle + \omega_4^2|\psi_2\rangle +\omega_4|\psi_3\rangle\right),
\end{eqnarray}
i. e.,
\begin{equation}
\label{eq145}
U_4 A_4 = A_4 U_4.
\end{equation}
It is clear that from Equation~\eqref{eq145}, we have 16 polynomials over 
the variables $\alpha_{k,l}$ for $0 \leq k, l \leq 3$. 
Since it is not easy to handle all these equations in hand, we use
Mathematica 7.0~\cite{math} to find the following conditions.
\begin{eqnarray*}
\alpha_{0,0} & = & \alpha_{2,2} + 2\alpha_{3,2} + 2\alpha_{0,3},\\
\alpha_{0,2} & = & \alpha_{2,2} + 2\alpha_{3,2} + \alpha_{0,3} - \alpha_{1,3} - 
\alpha_{2,3} - \alpha_{3,3},\\
\alpha_{2,0} & = & \alpha_{2,2} + \alpha_{0,3} - \alpha_{1,3} + \alpha_{2,3} - 
\alpha_{3,3},\\
\alpha_{1,0} & = & \alpha_{3,2} + \alpha_{0,3} - \alpha_{2,3},\\
\alpha_{3,0} & = & \alpha_{3,2} + \alpha_{0,3} - \alpha_{2,3},\\
\alpha_{0,1} & = & \alpha_{0,3},\\
\alpha_{1,1} & = & \alpha_{3,3},\\
\alpha_{3,1} & = & \alpha_{1,3},\\
\alpha_{2,1} & = & \alpha_{2,3},\\
\alpha_{1,2} & = & \alpha_{3,2}. 
\end{eqnarray*}
Hence the general form of 
$|\psi_0\rangle,|\psi_1\rangle, |\psi_2\rangle, |\psi_3\rangle$ such that 
it satisfies~(\ref{eqt144}) is as follows.
\begin{eqnarray}
\label{aaeq446}
|\psi_0\rangle & = & (\alpha_{2,2}+2\alpha_{3,2}+2\alpha_{0,3}) |0\rangle + \alpha_{0,3} |1\rangle \nonumber \\
& & + (\alpha_{2,2}+2\alpha_{3,2}+\alpha_{0,3}-\alpha_{1,3}-\alpha_{2,3}-\alpha_{3,3}) |2\rangle + \alpha_{0,3} |3\rangle \nonumber \\
|\psi_1\rangle & = & (\alpha_{3,2}+\alpha_{0,3}-\alpha_{2,3}) |0\rangle + \alpha_{3,3} |1\rangle + \alpha_{3,2} |2\rangle + \alpha_{1,3} |3\rangle \nonumber \\
|\psi_2\rangle & = & (\alpha_{2,2}+\alpha_{0,3}-\alpha_{1,3}+\alpha_{2,3}-\alpha_{3,3}) |0\rangle + \alpha_{2,3} |1\rangle + \alpha_{2,2} |2\rangle + \alpha_{2,3} |3\rangle \nonumber \\
|\psi_3\rangle & = & (\alpha_{3,2}+\alpha_{0,3}-\alpha_{2,3}) |0\rangle + \alpha_{1,3} |1\rangle + \alpha_{3,2} |2\rangle + \alpha_{3,3} |3\rangle
\end{eqnarray}
Thus we have the following result.
\begin{theorem}
Let $|\psi_0\rangle, |\psi_1\rangle, |\psi_2\rangle, |\psi_3\rangle$ be the
qudits as described in (\ref{aaeq44}). Then they will satisfy (\ref{eqt144})
if and only if they are of the form mentioned in (\ref{aaeq446}).
\end{theorem}
Similar to the previous sections, one may attempt to find out the conditions 
when $|\psi_0\rangle$, $|\psi_1\rangle$, $|\psi_2\rangle$, $|\psi_3\rangle$
are orthogonal. These cases as well as the cases for higher dimensions are not 
easy to handle by hand calculation and one may need to take the help of 
SAGE~\cite{sage} or Mathematica~\cite{math}.

\section{Conclusion and Open Directions}
\label{sec5}
In this paper we have studied the general classes of quantum states that 
can work in a similar manner as the standard bases with respect to the 
Quantum Fourier Transform. The QFT takes the state $|j\rangle$ 
to $\frac{1}{\sqrt{n}} \sum_{k=0}^{n-1} e^{\frac{2\pi i j k}{n}} |k\rangle$.
We have tried to characterize the states $|\psi_j\rangle$ that goes to
$\frac{1}{\sqrt{n}} \sum_{k=0}^{n-1} e^{\frac{2\pi i j k}{n}} |\psi_k\rangle$
under the action of QFT. We could provide a full characterization of the set 
of Hadamard-admissible pairs for qubits and as well for QFT-admissible 
triplets for qutrits. The generalized results for higher dimensions are also 
studied.


\begin{thebibliography}{0}

\bibitem{bb84}
C. H. Bennett and G. Brassard.
Quantum Cryptography: Public key distribution and coin tossing.
In Proceedings of the IEEE International Conference on Computers, 
Systems, and Signal Processing, Bangalore, India (IEEE, New York, 1984), 
pages 175--179.

\bibitem{bb384}
D. Bru\ss ~and C. Macchiavello.
Optimal eavesdropping in cryptography with three-dimensional quantum states.
Phys. Rev. Lett. 88 (2002) 127901 [quant-ph/0106126].

\bibitem{qDJ92}
\label{itx4}
D.~Deutsch and R.~Jozsa.
\newblock Rapid solution of problems by quantum computation,
\newblock {\em Proceedings of Royal Society of London}, A {\bf 439}, 553
(1992).

\bibitem{itx8}
A. Maitra and P. Parashar.
Hadamard Type Operations for Qubits, {\em International Journal of
Quantum Information}, volume 4, number 4, pages 653--664, August 2006.

\bibitem{math}
Mathematica 7.0.
Wolfram Research, Illinois Date of publication: 2008.

\bibitem{qNC02}
M.~A.~Nielsen and I.~L.~Chuang,
\newblock Quantum Computation and Quantum Information,
\newblock Cambridge University Press, 2002.

\bibitem{PT02}
A. K. Pati.
General impossible operations in quantum information.
{\em Physical Review A}, 66, 062319 (2002).

\bibitem{sage}
http://www.sagemath.org/

\bibitem{Sh94}
P. W. Shor.
\newblock Algorithms for quantum computation: discrete logarithms and
factoring,
\newblock {\em Proceedings of 35th Annual Symposium on Foundations of
Computer Science}, IEEE Press, Los Alamitos, CA (1994).

\bibitem{KY68}
K. Yosida, 
Functional Analysis, Springer Verlag, 1968.

\end{thebibliography}
\end{document}